# Multiplexed all-optical permutation operations using a reconfigurable diffractive optical network

Guangdong Ma[1,2,3,4,†], Xilin Yang[1,2,3,†], Bijie Bai[1,2,3], Jingxi Li[1,2,3], Yuhang Li[1,2,3], Tianyi Gan[1,3], Che-Yung Shen[1,2,3], Yijie Zhang[1,2,3], Yuzhu Li[1,2,3], Mona Jarrahi[1,3], Aydogan Ozcan[*,1,2,3,5]

[1]Electrical and Computer Engineering Department, University of California, Los Angeles, CA, 90095, USA.

[2]Bioengineering Department, University of California, Los Angeles, 90095, USA.

[3]California NanoSystems Institute (CNSI), University of California, Los Angeles, CA, 90095, USA.

[4]School of Physics, Xi'an Jiaotong University, Xi'an, Shaanxi, 710049, China.

[5]Department of Surgery, University of California, Los Angeles, CA, 90095, USA.

[†]These authors contributed equally

*Correspondence: Aydogan Ozcan, ozcan@ucla.edu

**Abstract**

Large-scale and high-dimensional permutation operations are important for various applications in e.g., telecommunications and encryption. Here, we demonstrate the use of all-optical diffractive computing to execute a set of high-dimensional permutation operations between an input and output field-of-view through layer rotations in a diffractive optical network. In this reconfigurable multiplexed material designed by deep learning, every diffractive layer has four orientations: $0°$, $90°$, $180°$, and $270°$. Each unique combination of these rotatable layers represents a distinct rotation state of the diffractive design tailored for a specific permutation operation. Therefore, a $K$-layer rotatable diffractive material is capable of all-optically performing up to $4^K$ independent permutation operations. The original input information can be decrypted by applying the specific inverse permutation matrix to output patterns, while applying other inverse operations will lead to loss of information. We demonstrated the feasibility of this reconfigurable multiplexed diffractive design by approximating 256 randomly selected permutation matrices using $K$=4 rotatable diffractive layers. We also experimentally validated this reconfigurable diffractive network using terahertz radiation and 3D-printed diffractive layers, providing a decent match to our numerical results. The presented rotation-multiplexed diffractive processor design is particularly useful due to its mechanical reconfigurability, offering multifunctional representation through a single fabrication process.

**Introduction**

Permutation operation[1] has been applied to a wide range of applications, including communications[2], data storage[3], remote sensing[4], and cryptography[5], among others;[6] it naturally conceals the original information by rearranging the positions of the individual input elements, thereby transforming the original message into a random-looking, unintelligible format[7,8]. There is a growing interest in the implementation of permutation operations in an all-optical manner, as opposed to relying on specialized electronic hardware. This transition is partially motivated by the aspiration to improve the efficiency and scalability of information processing through the seamless integration of optical methods. Therefore, considerable research has been dedicated to the development of all-optical permutation operations by leveraging the intrinsic properties of light with specialized optical materials and components. Examples of these efforts include the utilization of multiwavelength photonic crystal ring resonators[9], polarization-based coherent optical communication systems[10], and cascaded Fourier transformation systems[11]. Such optical permutation operations act as data shuffling structures capable of sorting the input information into the desired order. Moreover, these optical systems offer advantages in information transfer and processing with their power efficiency and ultrahigh-speed, making them promising for applications in optical switching[9,12] and encryption[10,11]. Nevertheless, existing approaches face limitations in terms of parallelism and scalability for applications in high-dimensional optical communications due to their limited degrees of freedom and throughput.

Here, we demonstrate a reconfigurable multiplexed diffractive material to all-optically perform a large set of high-dimensional permutation operations using a single diffractive architecture that is based on layer rotations. Diffractive deep neural networks ($D^2$NNs) composed of spatially-engineered diffractive layers emerged as a compelling solution to perform high-dimensional and high-throughput analog processing of optical information[13–16]. A diffractive optical network processes the wavefront of the input light field as it propagates through a series of structured diffractive surfaces connected via free-space propagation through a thin volume, enabling massive parallelism and high speed for various advanced optical information processing tasks[17–23], including image classification[24–31], quantitative phase imaging (QPI)[32–34], imaging through random diffusers[35,36], single-pixel imaging[37], multispectral imaging[38], super-resolution displays[39], logic operations[40,41], among others[42–48]. This work reports a reconfigurable multiplexed $D^2NN$ architecture, termed $R\text{-}D^2NN$, using rotatable diffractive layers that collectively approximate a large set of independent permutation operations through the same optimized diffractive layers. Within this $R\text{-}D^2NN$ design, each rotatable diffractive layer offers four pre-determined orientations: $0°$, $90°$, $180°$, and $270°$. Each unique combination of these rotatable diffractive layers of an $R\text{-}D^2NN$ design represents a distinct diffractive processor, executing one specific task, i.e., a given permutation operation. Therefore, $K$ rotatable diffractive layers of an $R\text{-}D^2NN$ would have in total $4^K$ rotation states, enabling all-optical execution of up to $4^K$ independent linear operations; for each one of these independent operations, infinitely many distinct input patterns will be accurately transformed following the target transformation matrix. We demonstrate the use of $R$ rotation states of an $R\text{-}D^2NN$ design to perform $R$ independent/unique permutation operations between its input and output fields-of-views (FOVs) using a $K$-layer diffractive material (i.e., $R \leq 4^K$). After employing a predefined rotation architecture, the rotatable diffractive layers of the $R\text{-}D^2NN$ are optimized by a stochastic gradient-descent-based learning algorithm. This deep learning-based optimization aims to accurately execute a distinct preassigned permutation operation on every input image within its input FOV, ultimately producing a permutated or encrypted image at its output FOV – performing a unique permutation operation assigned to a rotation state of the reconfigurable diffractive material. Our analyses demonstrate that when the total number ($N$) of trainable diffractive features approaches $\sim RN_iN_o$, $R\text{-}D^2NN$ architecture can successfully implement $R$ unique permutation operations with negligible error using pre-determined rotations of the same diffractive layers, where $N_i$ and $N_o$ represent the number of diffraction-limited pixels at the input and output FOVs, respectively. Based on this multiplexed architecture, we report the accurate approximation of 256 unique permutation matrices using $K = 4$ rotatable diffractive layers ($R = 4^K = 256$). We also experimentally confirm the success of the R-D$^2$NN framework using terahertz radiation and a 3D-printed reconfigurable diffractive processor, providing a decent agreement with our numerical results. The presented framework of mechanically reconfigurable multiplexed diffractive optical materials holds promise for inspiring resource-efficient physical designs in data encryption, computational imaging, sensing, and high-dimensional optical communication.

**Results**

**Design and training of R-D$^2$NN architecture**

Figure 1 (a) illustrates the schematic of the R-D$^2$NN design under an illumination wavelength of $\lambda$. This design encompasses $K$ rotatable layers, each of which has four independent orientations: 0°, 90°, 180°, and 270°. Each unique orientation combination of the entire $K$ rotatable layers corresponds to a *rotation state* of the R-D$^2$NN, and we also refer to each rotation state of R-D$^2$NN as a unique D$^2$NN configuration. Hence, a $K$-layer rotatable R-D$^2$NN system can potentially permit $4^K$ different rotation states, or D$^2$NN configurations, performing up to $4^K$ transformations if each rotation state is preassigned to an arbitrary linear transformation. In the design phase, our framework is iteratively trained to approximate $R$ unique permutation matrices between the input and output FOVs with $R$ distinct rotation states of the R-D$^2$NN ($R \leq 4^K$). The selected orientations for $R$ rotation states constitute one *rotation architecture* of the R-D$^2$NN. The axial distance between the successive planes of the R-D$^2$NN is set as $d$ (Figure 1a), and the free space propagation between the successive planes is modeled by the angular spectrum method[42].

The output intensity of R-D$^2$NN for each rotation state $i$ is denoted as $\hat{O}_i = \hat{P}_i G$ ($i = 1, 2, \ldots R$), where $\hat{P}_i \in \mathbb{R}^{N_i * N_o}$ signifies the permutation operation executed by R-D$^2$NN in the rotation state $i$. Here, $N_i$ and $N_o$ represent the number of pixels at the input and output FOVs, respectively. During the decryption operation, applying the specific inverse permutation matrix $P_j^{-1}$ ($j = i$) enables the recovery of the original input image intensity ($G$), while applying unmatched inverse operation $P_j^{-1}$ ($j \neq i$) leads to a speckle-like, concealed pattern. Figure 1 (b) illustrates the permutation matrices preassigned to each rotation state. Each permutation matrix, $P_i$, acts as a distinctive mapping that transforms the intensity of an image from the input plane ($G$) to the output plane ($O_i = P_i G$). The permutated image $P_i G$ is used as the ground truth of the R-D²NN in the rotation state $i$ during the supervised training and blind testing/evaluation phases.

To optimize the R-D²NN, we created a customized dataset containing 60,000 random image data (see the Methods section for details) and minimized the normalized mean-squared error (NMSE) between the ground truth $O_i = P_i G$ and the output intensity pattern $\hat{O}_i = \hat{P}_i G$. The physical output of a R-D²NN in rotation state $i$ can be written as:

$$\hat{P}_i G = \hat{O}_i = \left| \hat{A}_i \sqrt{G} \right|^2 \qquad (1).$$

where $\hat{A}_i$ denotes the matrix description of the complex-valued forward model operator of the R-D²NN in rotation state $i$. There are $N$ trainable diffractive neurons/features evenly spaced on the $K$ rotatable layers of an R-D²NN design, i.e., $\frac{N}{K}$ trainable diffractive features on each rotatable layer, collectively modulating the phase of the incident wavefront. These diffractive features (each with a lateral size of ~0.53$\lambda$) on each layer are optimized using a stochastic gradient-descent-based algorithm to minimize the NMSE-based loss function. More details about the R-D²NN, optical forward model, and training details can be found in the Methods part.

**Performance analysis of different rotation architectures**

Following the schematic shown in Figure 1, we illustrate a layout of R-D$^2$NN featuring $K = 4$ rotatable diffractive layers with a total number of $1.5RN_iN_o$ diffractive features in Figure 2a ($R = 4, N_i = N_o = 64$). In this design, we randomly select four rotation states from all possible ($4^4$) states of R-D$^2$NN to approximate four distinct permutation matrices. This R-D$^2$NN design encrypts the original image through different rotation states, resulting in non-informative measurement patterns at the output plane. In this design, we utilized a total number of $1.5RN_iN_o$ diffractive features, and the rotation architecture is shown in Figure 2 (f), which was randomly selected. After deep learning-based optimization, the phase masks of the resulting R-D$^2$NN are shown in Figure 2b. To evaluate each estimated permutation operation corresponding to each rotation state of the R-D$^2$NN, we numerically tested the spatially varying point spread functions of the optimized R-D$^2$NN, revealing the estimated permutation matrix $\hat{P}_i$ implemented by the diffractive processor in the rotation state $i$. Figures 2c-d depict the four preassigned/target permutation matrices $P_i$ ($i = 1,2,3,4$) and optically estimated permutation matrices $\hat{P}_i$ ($i = 1,2,3,4$), as well as the difference ($P_i$-$\hat{P}_i$) between them, which provide a very good agreement with each other. To quantitatively assess the performance of the R-D$^2$NN, we calculated two additional metrics: 1) cosine similarity, and 2) NMSE between $\hat{P}_i$ and $P_i$. As shown in Figure 2 (e), the cosine similarity between $P_i$ and $\hat{P}_i$ is approximately one, and the NMSE between them is around $3 \times 10^{-6}$, both of which indicate that the all-optical permutation operations executed by R-D$^2$NN match their target permutation operations very well. The R-D$^2$NN successfully learns the permutation operation with negligible error and demonstrates accurate performance on any given input intensity pattern, despite being trained on a customized random image dataset. Remarkably, the R-D$^2$NN exhibits exceptional generalization capabilities when testing with diverse datasets never included in the training phase; see e.g. Figures 3 (a1) and (b1) for two examples. In the decryption stage, four distinct inverse permutation operations are applied to the results of the diffractive processor in each rotation state, and the decryption results are presented in a confusion matrix-like format. A faithful reconstruction of the original image is achievable only by employing the corresponding inverse permutation operation $P_j^{-1}$ $(j = i)$, as illustrated by the diagonal elements. In contrast, the application of any other inverse permutation matrix $P_j^{-1}$ $(j \neq i)$ results in noise-like images, as depicted in the off-diagonal elements. As shown in Figures 3 (a2), (b2) (a3), and (b3), we also calculated two additional quantitative metrics, 1) the Pearson correlation coefficient (PCC) value between the decrypted image ($\hat{G} = P_i^{-1}\hat{P}_i G$) and the original image ($G$), and 2) the diffraction efficiency of each D$^2$NN configuration across all 20,000 blind testing image data from different datasets. The diagonal PCC values consistently approach $\sim$1 for the decryption results regardless of which dataset is tested, demonstrating the success of our R-D$^2$NN design in all-optically encrypting images. The diffraction efficiency values shown in Fig. 3 are obtained from the R-D$^2$NN model that was trained without any efficiency penalty. A higher diffraction efficiency R-D$^2$NN model can be obtained by introducing an additional energy penalty term in the loss function (see the Methods part).

In this design scheme, the 4-layer R-D$^2$NN can potentially offer up to $4^4$ independent rotation states. If we randomly select four rotation states out of all possible states, it gives rise to a total of $C(^{256}_4)$ possible rotation architectures. To understand whether employing different rotation architectures has an impact on the performance of the 4-layer R-D$^2$NN design, we randomly selected 10 rotation architectures from $C(^{256}_4)$ possible combinations; see Figure 4 (c). Each of these 10 rotation architectures ($A_i, i = 1, 2, \ldots 10$) is preassigned to the same permutation matrices shown in Figure 2 (c) to provide a fair comparison of their performance. In the same analysis, we also investigated the performance of the 4-layer R-D$^2$NN designs as a function of the total number of trainable diffractive features/neurons; for this, each one of the randomly selected rotation architectures was trained with $N = 1.0 R N_i N_o$, $1.5 R N_i N_o$, and $2.0 R N_i N_o$ trainable diffractive features. After their deep learning-based optimization, the performances of these R-D$^2$NN models with various rotation architectures and varying numbers of trainable features are summarized in Figures 4 (a) and (b). We evaluated their performance using two metrics: 1) the cosine similarity, and 2) the NMSE between the estimated ($\hat{P}_i$) and the target permutation matrices ($P_i$). Our analyses revealed that there was no statistically significant variation in the cosine similarity and NMSE values of these different rotation architectures and that a further increase in the number of trainable features ($N$) improved the approximation accuracy of these R-D$^2$NN designs as expected.

**Performance analysis as a function of the number of rotatable diffractive layers**

To further investigate the performance of R-D$^2$NN designs with varying numbers of rotatable diffractive layers, we kept the total number of trainable diffractive features/neurons the same, while changing the number of rotatable diffractive layers; see Figure 5. In this comparison, we kept the number of rotations equal to 4 ($R = 4$) and adopted the first rotation architecture shown in Figure 4(c) across all the R-D$^2$NN models with $N_i = N_o = 64$. Figure 5 (a) illustrates an example from a set of 10,000 blind testing results, showcasing a handwritten digit '5' as the input image of R-D$^2$NN. Figures 5(b)-(e) exhibit the results of four different R-D$^2$NN models featuring 2, 3, 4, and 5 rotatable diffractive layers, while maintaining $N = 1.0RN_iN_o$. Each output intensity is decrypted by applying inverse permutation matrices $P_j^{-1}$ as shown on the right side of the dashed line in Figure 5, where only the specific inverse permutation matrix ($j = i$) can successfully decrypt the input image. The image decryption results of a 5-layer design achieved a PCC value of 0.98 and a PSNR (Peak Signal-to-Noise Ratio) value of 22 dB, outperforming shallower designs with 2-, 3-, or 4-layers.

We also trained R-D$^2$NN designs with different total numbers of diffractive features/neurons covering $N = 1.0RN_iN_o$, $N = 1.5RN_iN_o$, and $N = 2.0RN_iN_o$, spread over $K$ = 2, 3, 4, and 5 rotatable diffractive layers. Figure 6 (a) illustrates the cosine similarity between the desired permutation operations ($P_i$) and their all-optical versions ($\hat{P}_i$) as a function of $K$ and $N$, while Figure 6 (b) depicts the NMSE between them. The quantitative comparisons reported in Figure 6 suggest two ways to enhance the performance of R-D$^2$NN designs: 1) increasing $N$; and 2) distributing the total number of diffractive neurons across a deeper diffractive architecture, in line with earlier results on diffractive processors[13,26,45,51].

**Scaling the number of rotation states in R-D$^2$NN**

Thus far, we have demonstrated that R-D$^2$NN can be designed to approximate $R$ = 4 arbitrary permutation matrices. Next, we explore the scalability of R-D$^2$NN in approximating a larger number of permutation operations by involving more rotation states in its design. To investigate the performance of R-D$^2$NN designs as a function of $R$, we considered different numbers of rotations $R \in \{4, 8, 16, 32, 64, 128, 196\}$. In this comparative analysis, we utilized $N = 1.5 R N_i N_o$ with $N_i = N_o = 5 \times 5$ for each architecture. We used a new set of 196 different arbitrarily selected permutation matrices, each with a dimension of $25 \times 25$. Training, validation, and testing datasets are created using the same approach as described earlier.

Figures 7(a)-(g) depict the NMSE between the output ($\hat{O}_i$) of the diffractive processor in rotation state $i$ and the ground truth ($O_i$) as a function of $R$ ($R \in \{4, 8, 16, 32, 64, 128, 196\}$). As expected, we observe that the all-optical diffractive processors exhibit increasing NMSE values as $R$ increases from $R = 4$ (Figure 7(a)) to ($R = 196$) (Figure 7 (g)). To further shed light on this analysis, we also trained R-D$^2$NN designs with $N = 1.0 R N_i N_o$, $N = 1.5 R N_i N_o$, and $N = 2.0 R N_i N_o$; Figures 8 (a) and (b) show the cosine similarity and the NMSE between $P_i$ and their all-optical versions ($\hat{P}_i$), where the mean and the standard deviation values were calculated across $R$ permutation matrices in each case. Figure 8 (c) also presents the NMSE between the output of the R-D$^2$NN designs and the ground truth images, calculated across 10,000 blind-testing images. Linearly extrapolating the data points reported in Figures 8 (b) and 8(c) allows us to estimate the expected performance at $R = 256$, which is challenging to simulate due to limited computational resources. Based on this extrapolation, a rough estimation of the PSNR between the R-D$^2$NN output and the ground truth images at $N_R = 256$ rotations yields 17.8dB, 21.8dB, and 25.4dB for $N = 1.0 R N_i N_o$, $N = 1.5 R N_i N_o$, and $N = 2.0 R N_i N_o$, respectively. These analyses further support that increasing the number of trainable diffractive features in an R-D²NN design significantly enhances its performance, particularly in cases with a larger number of rotation states.

**Comparison between layer rotation-based multiplexing and wavelength-based multiplexing**

To compare the performance of different multiplexing methods, we implemented a 4-layer wavelength-based multiplexing scheme[46] to create a broadband diffractive optical processor with four different wavelengths, $\lambda_1 = 0.70mm$, $\lambda_2 = 0.73mm$, $\lambda_3 = 0.76mm$, $\lambda_4 = 0.80mm$ – each wavelength assigned to approximate one permutation matrix. The mean wavelength of this group, $\lambda_m = 0.75mm$, is employed for the rotation-based multiplexing method. We selected the first rotation architecture of Figure 4(c) for the four rotation states within a 4-layer R-D²NN. In this comparison, four preassigned target permutation matrices (Figure 2 c) for two multiplexing methods are kept the same. Other parameters, such as the lost function, datasets, learning rate, material dispersion, and layer-to-layer distance, also remain the same to provide a fair comparison.

Figures S2 (a1) and (b1) depict the cosine similarity between the desired permutation matrices ($P_i, i = 1, 2, 3, 4$) and their all-optical counterparts ($\hat{P}_i, i = 1, 2, 3, 4$) as a function of the total number of trainable neurons/features for the two multiplexing methods. These results indicate that an increased number of trainable diffractive features within each multiplexing method significantly enhances the output performance. Interestingly, the rotation-based multiplexing method outperforms the wavelength-based multiplexing method when using the same number of diffractive features (see Figure S2). Moreover, the cosine similarity between the desired permutation operations ($P_i$) and the estimated permutation operations ($\hat{P}_i$) of the rotation-based method with $N = 1.5RN_iN_o$ is higher than that of the wavelength-based multiplexing method with $N = 2.0RN_iN_o$ as shown in Figure S2 (a2). The same conclusion can also be drawn from the NMSE results reported in Figure S2 (b2).

**Experimental validation**

We performed a proof-of-concept experimental validation of a 3-layer R-D$^2$NN design by fabricating and assembling the diffractive layers using a 3D printer and illuminating them with a terahertz (THz) source at $\lambda = 0.75$mm (Figure 9a). As shown in Figure 9 (b), we designed this diffractive processor with a vaccination strategy[28] during the training process by introducing random displacements to the diffractive layers taking into account potential experimental misalignments and fabrication artifacts (see the Methods section for details). Our prototype was designed with two rotation states $R = 2$ and $N_i = N_o = 5^2$. Each rotation state in the experiment is assigned one permutation matrix, such that any spatially structured pattern at the input FOV can be all-optically permutated by the R-D$^2$NN into the different desired patterns at the output FOV. To do this, the first and the second diffractive layers are kept at $0°$, while the third layer is rotated to $0°$ or $180°$ to perform two permutation matrices $P_1$ and $P_2$, respectively. During the training process, a set of 40,000 randomly generated input-output pairs corresponding to the target permutation matrices ($P_1$ and $P_2$) were used to optimize the thickness values of diffractive layers. An additional energy efficiency penalty loss was added to the training loss function for sufficient output power to be detected in the experiments (see the Methods section for details). After the R-D$^2$NN converged, the resulting diffractive layers were fabricated using a 3D printer and mechanically assembled, forming a physical rotatable diffractive optical permutation processor as shown in Figure 9 (c).

To experimentally test the performance of this 3D-fabricated R-D$^2$NN, the permutated versions of two letters 'U', and 'C' were prepared as the input test objects. Each letter is permutated by the inverse permutation matrices $P_1^{-1}$ and $P_2^{-1}$, respectively, as shown in Figures 10 (a1) and (b1). The experimental results are presented in Figures 10 (c1) and (c2), demonstrating a decent agreement with the numerical simulation results shown in Figures 10 (b1) and (b2). Only the diagonal entries/figures deliver readable information, confirming the feasibility of our R-D$^2$NN design.

**Discussion**

In this work, we introduced a mechanically reconfigurable diffractive network with diffractive layer rotations to perform a large set of permutation operations. In this design, passive diffractive layers execute the permutation/encryption on the input optical fields, providing an energy-efficiency approach without requiring any external power source and completing its computational task at the speed of light. Our findings confirm the robustness of the rotation-based multiplexing method across various randomly selected rotation architectures, particularly when a sufficient number of trainable diffractive features/neurons are available. We also quantified the scalability of the rotational states of the R-$D^2$NN designs and compared their performance with a wavelength-based multiplexing method, highlighting the superior efficacy of the layer-rotatable designs. Furthermore, our proof-of-concept experimental results verified the feasibility of our mechanically reconfigurable diffractive designs. The presented R-$D^2$NN framework enhances the versatility of all-optical diffractive computing systems by introducing an additional degree of freedom alongside wavelength multiplexing[46] and polarization multiplexing[16,47].

In real-world applications, the axial and lateral misalignments of diffractive layers may degrade the performance of R-$D^2$NN, especially if rapid switching between different rotation states is needed. To mitigate this issue, one strategy is to introduce random errors into the physical forward model, a process referred to as 'vaccination' of the diffractive network, during the training of an experimental design[28]. This process involves adding random displacements in the lateral and axial positions of the diffractive layers, which helps build resilience against potential misalignments of the physical system. Additional factors, such as absorption, surface reflections of the material, and fabrication-related imperfections, could also degrade the performance of R-$D^2$NN designs, which might be potentially mitigated using high-precision lithography and antireflection coatings during the fabrication process.

Another practical limitation could be the relatively lower diffraction efficiency of R-D$^2$NN designs, potentially providing insufficient signal-to-noise ratio (SNR) for certain applications. Nevertheless, the diffraction efficiency of a diffractive processor can be significantly enhanced by including additional loss terms to penalize the poor energy efficiency of the network, with a minimal compromise in performance[48]. In addition, increasing the total number of trainable diffractive features, i.e., $N \gg RN_iN_o$, could also improve the diffraction efficiency, especially in the case of a vaccinated diffractive network design[14].

Finally, while the decryption process of the reported mechanically reconfigurable R-D$^2$NNs was implemented digitally, it could also be replaced by another jointly optimized R-D$^2$NN model trained to all-optically perform the desired inverse permutation operations for each rotation state of the diffractive processor. Such an all-optical approach might enable the integration of encryption and decryption R-D$^2$NNs at both sides of a transmission link, achieving mechanically reconfigurable switching and permutation operations, which can also be expanded to other linear transformations of interest.

**Methods**

**Forward model of layer-rotatable multiplexed diffractive neural networks**

For a $K$-layer rotatable R-D$^2$NN, the forward propagation of the complex field can be modeled as a sequence of free-space propagation between the $l$th and ($l$+1)th layers ($l = 0, 1, \ldots, K$), and the modulation of the $l$th diffractive layer/mask ($l = 1, \ldots, K$), where 0th layer is the input plane and ($K + 1$)th layer is the output plane. The complex field $u^l(x, y)$ propagates from $l$th to $(l + 1)$th layer and is calculated by the angular spectrum method[40]:

$$u^{l+1}(x,y) = \mathbb{P}_d u^l(x,y) = \mathcal{F}^{-1}\{\mathcal{F}\{u^l(x,y)\}H(f_x, f_y, d)\} \quad (1)$$

where $\mathbb{P}_d$ is the free-space propagator, and $d$ represents the axial distance between two successive planes. $\mathcal{F}$ and $\mathcal{F}^{-1}$ denote the Fourier transform and the inverse Fourier transform operations, respectively. The transfer function $H(f_x, f_y, d)$ is defined as:

$$H(f_x, f_y, d) = \begin{cases} \exp(j2\pi d\sqrt{1/\lambda^2 - f_x^2 - f_y^2}, & 1/\lambda^2 - f_x^2 - f_y^2 > 0 \\ 0, & 1/\lambda^2 - f_x^2 - f_y^2 \leq 0 \end{cases} \quad (2)$$

where $j = \sqrt{-1}$, $k = \frac{2\pi}{\lambda}$, and $\lambda$ is the wavelength of the input field. $f_x$ and $f_y$ denote the spatial frequencies. The $l$th diffractive layer/mask modulates the phase of the transmitted field with the transmittance coefficient $t^l(x, y)$:

$$t^l(x, y) = \exp\left(j\phi^l(x, y)\right) \quad (3)$$

where $\phi^l(x, y)$ denotes the phase modulation of the learnable diffractive feature/neuron located at the $l$th diffractive mask. Based on Eqs. (1)-(3), the transmitted complexed field propagated through the $K$th layer can be expressed as:

$$u^K(x, y) = \left(\prod_{l=1}^{K} t^l(x, y) \, \mathbb{P}_d\right) u^0(x, y) \quad (4).$$

After modulation by all the $K$ diffractive layers, the resulting light field is further propagated an axial distance of $d$ to the output plane, and the output intensity $\hat{O}_i(x, y)$ of the R-D²NN in rotation state $i$ can be written as:

$$\hat{O}_i(x, y) = |u^{K+1}(x, y)|^2 = |\mathbb{P}_d u^K(x, y)|^2$$

**Random selection of orthogonal permutation matrices**

In this paper, the input and output FOVs of the diffractive networks are established with an identical size of $5 \times 5$, or $8 \times 8$ pixels, i.e., $N_i = N_o \in \mathbb{R}^{5\times 5}$, or $N_i = N_o \in \mathbb{R}^{8\times 8}$. As a result, the size of the permutation matrix $P_i$ is equal to $25 \times 25$, or $64 \times 64$, respectively, according to the input and output FOVs. We prepared a randomly selected set of orthogonal permutation matrices $P_i$ ($i = 1, 2, \ldots R$) and subsequently assigned them to $R$ selected rotation states of a R-D²NN design for all-optical encryption. For generating a large set of orthogonal permutation matrices, we used a block-permutation approach to ensure orthogonality between two matrices:

$$\sum_{i,j} a_{i,j} * b_{i,j} = \sum_{n}\sum_{i,j} a_{i,j}^n * b_{i,j}^n = 0 \tag{6}$$

where $a_{i,j}$ and $b_{i,j}$ denote elements of two permutation matrices, $n$ denotes the number of blocks within each set. We ensured that the current block was randomly permuted while satisfying orthogonality in the previous blocks. This way, the block-permutation algorithm guarantees the orthogonality of corresponding blocks among all permutation matrices, as elucidated in Figure S1(a-b). As shown in Figure S1(c-d), this approach efficiently generates a large set of orthogonal permutation matrices with no inter-correlation.

**Network training**

The $K$-layer rotatable R-D²NNs in this study are designed for monochromatic coherent illumination with a wavelength of $\lambda$ ($\lambda = 0.75$mm). Each rotatable diffractive layer/mask contains $\frac{N}{K}$ trainable neurons with a pixel/neuron size of $0.53\lambda$ (0.40mm), allowing the phase modulation of the transmitted field, where $N$ denotes the total number of trainable neurons/features within the R-D²NN. To maintain the full connectivity between any two successive planes[13], the axis distance between any two successive planes of the R-D²NN is set as $0.3\sqrt{\frac{N}{K}}\lambda$, i.e., $d_{l,l+1} = 0.3\sqrt{\frac{N}{K}}\lambda$ ($l = 0,1,\dots K$). During the training process of the R-D²NN, each $8 \times 8$-pixel or $5 \times 5$-pixel random raw image is first up-sampled by a factor of 4 to match the input FOVs of the R-D²NN design. The permutated version of the random input, $P_iG$, serves as the ground truth of the R-D²NN, where $P_i$ means the permutation operation preassigned to the R-D²NN in rotation state $i$. The encrypted result of the R-D²NN in rotation state $i$, i.e., $\hat{P}_iG$, has the same dimension as the input random image, where $\hat{P}_i$ means the permutation operation executed by the R-D²NN in rotation state $i$. According to the input and output FOVs ($N_i = N_o \in \mathbb{R}^{5\times5}$, or $N_i = N_o \in \mathbb{R}^{8\times8}$), each permutation matrix $P_i$ has the dimension of $64 \times 64$-pixel or $25 \times 25$-pixel, digitally generated using a block-permutated algorithm as described in the "Random selection of orthogonal permutation matrices" section.

We created a customized dataset containing 60,000 randomly generated image data, where the value of each pixel is randomly distributed between 0 and 1. This random image dataset is divided into training, validation, and testing datasets without any overlap. Each dataset contains 40,000, 10,000, and 10,000 random images, respectively. To investigate the generalization capabilities of the optimized R-D$^2$NN model, we used 10,000 images of the MNIST handwritten dataset during the blind testing process. The R-D$^2$NN is trained through the Adam optimizer[50] with a learning rate of 0.01 up to 50 epochs for all tasks. All the models were trained and tested using PyTorch1.12[51] with a GeForce RTX 1080 graphical processing unit (NVIDIA Inc.). The typical time required for training a R-D$^2$NN with $K = 4$, $R = 4$, $N_i = N_o = 8 \times 8$, and $N = 1.5RN_iN_o$ is ~0.5h.

**Experimental design**

For the experimentally validated rotatable diffractive design, an energy efficiency penalty is used by setting the constant $\alpha_2 = 1$ in Eq. (7). During the training, we used a random image dataset with a size of $50,000 \times 5 \times 5$, 10,000 of which are selected as the validation dataset. Each diffractive layer contains $120 \times 120$ trainable diffractive neurons/features, each with a size of $0.53\lambda$ (0.40mm). We vaccinated the experimental R-D$^2$NN model during the training by introducing random displacements in the lateral direction for the rotatable diffractive layer and the axial direction for the free-space propagation model between successive planes, respectively[28]. Specifically, a randomly generated 3D displacement vector $D = (D_x, D_y, D_z)$ was added to each layer and free-space propagation:

$$D_x = \mathrm{U}(-\Delta_{x,tr}, \Delta_{x,tr}) \tag{14}$$

$$D_y = \mathrm{U}(-\Delta_{y,tr}, \Delta_{y,tr}) \tag{15}$$

$$D_z = \text{U}(-\Delta_{z,tr}, \Delta_{z,tr}) \qquad (16)$$

where U represents a uniform random distribution. $D_x$ and $D_y$ denote the vaccination for diffractive layers in the lateral direction, while $D_z$ denotes the random vaccination for the free-space propagation model between successive planes. $\Delta_{*,tr}$ represents the maximum amount of the shift along the corresponding axis throughout the training process; during the training we set $\Delta_{x,tr} = \Delta_{y,tr} = 1.13$mm ($\sim 1.5\lambda$), and $\Delta_{z,tr} = 0.75$mm ($\sim 1.0\lambda$).

**Terahertz experimental setup**

The phase values of each diffractive layer are converted into height maps $h^l(x, y)$ using the equation $h^l(x, y) = \phi^l(x, y)\frac{\lambda}{2\pi(n-1)}$, where $\phi^l(x, y)$ refers to the optimized phase values corresponding to the $l$th layer. $n$ is the refractive index of the printed material (VeroBlackPlus RGD875). We fabricated diffractive layers and tested objects using a 3D printer (Object30 V5 Pro, Stratasys). We also used a custom-designed holder printed by a 3D printer (Object30 Pro, Stratasys) to assemble the printed diffractive layers along with the input objects.

We validated the fabricated R-D$^2$NN design using a THz continuous wave scanning system as illustrated in Figure 9(a). The THz source consists of a modular amplifier (Virginia Diode Inc. WR9.0M SGX)/multiplier chain (Virginia Diode Inc. WR4.3x2 WR2.2x2) (AMC), followed by a compatible diagonal horn antenna (Virginia Diodes Inc. WR 2.2). The input of AMC was a 10 dBm RF signal at 11.1111 GHz ($f_{RF1}$) and after being multiplied 36 times, the output radiation was at 0.4 THz. The AMC was modulated with a 1 kHz square wave for lock-in detection. The output plane of the diffractive design was scanned with a step size of 1.6mm using a mixer (Virginia Diode Inc. WRI 2.2). The mixer is mounted on an XY positioning stage that was built by combining two linear motorized states (Thorlabs NRT100). A 10 dBm RF signal at 11.0833 GHz ($f_{RF2}$) was sent to the detector as a local oscillator to down-convert the signal to 1 GHz. The down-converted signal was amplified by a low-noise amplifier (Mini-Circuits ZRL-1150-LN+) and filtered by a 1 GHz ($\pm 10\ MHz$) bandpass filter (KL Electronic 3C40-1000/T10-O/O). Then, the signal was read by a low-noise power detector (Mini-Circuits ZX47-60) for absolute power detection. The detector output was measured by a lock-in amplifier (Stanford Research SR830) with the 1 kHz square wave used as the reference signal. Then, the lock-in amplifier readings were calibrated into a linear scale. A digital $4 \times 4$ binning was applied to each measurement to match the training feature size of the design phase.

**Supporting Information**: This file contains:

- Supplementary Figures S1-S2.
- Training loss function
- Performance evaluation metrics

**References**

[1] D. Slepian, *Proc. IEEE* **1965**, *53*, 228.
[2] R. F. Siregar, N. Rajatheva, M. Latva-Aho, *IEEE Trans. Wirel. Commun.* **2023**, *22*, 4947.
[3] Anxiao Jiang, R. Mateescu, M. Schwartz, J. Bruck, *IEEE Trans. Inf. Theory* **2009**, *55*, 2659.
[4] X. Huang, G. Ye, H. Chai, O. Xie, *Secur. Commun. Netw.* **2015**, *8*, 3659.
[5] S. G. Akl, H. Meijer, in *Adv. Cryptol.* (Eds.: G. R. Blakley, D. Chaum), Springer, Berlin, Heidelberg, **1985**, pp. 269–275.
[6] N. Ishikawa, S. Sugiura, L. Hanzo, *IEEE Commun. Surv. Tutor.* **2018**, *20*, 1905.
[7] H. Huang, X. He, Y. Xiang, W. Wen, Y. Zhang, *Signal Process.* **2018**, *150*, 183.
[8] R. Enayatifar, A. H. Abdullah, I. F. Isnin, A. Altameem, M. Lee, *Opt. Lasers Eng.* **2017**, *90*, 146.
[9] M. Djavid, M. H. T. Dastjerdi, M. R. Philip, D. D. Choudhary, T. T. Pham, A. Khreishah, H. P. T. Nguyen, *Photonic Netw. Commun.* **2018**, *35*, 90.
[10] S. Ishimura, K. Kikuchi, *Opt. Express* **2015**, *23*, 15587.
[11] M. He, Q. Tan, L. Cao, Q. He, G. Jin, *Opt. Express* **2009**, *17*, 22462.
[12] R. A. Spanke, V. E. Benes, *Appl. Opt.* **1987**, *26*, 1226.
[13] X. Lin, Y. Rivenson, N. T. Yardimci, M. Veli, Y. Luo, M. Jarrahi, A. Ozcan, *Science* **2018**, *361*, 1004.

[14] D. Mengu, Y. Zhao, A. Tabassum, M. Jarrahi, A. Ozcan, *Nanophotonics* **2023**, *12*, 905.
[15] B. Bai, H. Wei, X. Yang, T. Gan, D. Mengu, M. Jarrahi, A. Ozcan, *Adv. Mater.* **2023**, *35*, 2212091.
[16] Y. Li, J. Li, Y. Zhao, T. Gan, J. Hu, M. Jarrahi, A. Ozcan, *Adv. Mater.* **n.d.**, *n/a*, 2303395.
[17] Y. Shen, N. C. Harris, S. Skirlo, M. Prabhu, T. Baehr-Jones, M. Hochberg, X. Sun, S. Zhao, H. Larochelle, D. Englund, M. Soljačić, *Nat. Photonics* **2017**, *11*, 441.
[18] G. Wetzstein, A. Ozcan, S. Gigan, S. Fan, D. Englund, M. Soljačić, C. Denz, D. A. B. Miller, D. Psaltis, *Nature* **2020**, *588*, 39.
[19] H. Zhang, M. Gu, X. D. Jiang, J. Thompson, H. Cai, S. Paesani, R. Santagati, A. Laing, Y. Zhang, M. H. Yung, Y. Z. Shi, F. K. Muhammad, G. Q. Lo, X. S. Luo, B. Dong, D. L. Kwong, L. C. Kwek, A. Q. Liu, *Nat. Commun.* **2021**, *12*, 457.
[20] U. Teğin, M. Yıldırım, İ. Oğuz, C. Moser, D. Psaltis, *Nat. Comput. Sci.* **2021**, *1*, 542.
[21] T. Zhou, X. Lin, J. Wu, Y. Chen, H. Xie, Y. Li, J. Fan, H. Wu, L. Fang, Q. Dai, *Nat. Photonics* **2021**, *15*, 367.
[22] F. Ashtiani, A. J. Geers, F. Aflatouni, *Nature* **2022**, *606*, 501.
[23] "Experimentally realized in situ backpropagation for deep learning in photonic neural networks | Science," can be found under https://www.science.org/doi/10.1126/science.ade8450, **n.d.**
[24] J. Li, D. Mengu, Y. Luo, Y. Rivenson, A. Ozcan, *Adv. Photonics* **2019**, *1*, 046001.
[25] T. Yan, J. Wu, T. Zhou, H. Xie, F. Xu, J. Fan, L. Fang, X. Lin, Q. Dai, *Phys. Rev. Lett.* **2019**, *123*, 023901.
[26] D. Mengu, Y. Luo, Y. Rivenson, A. Ozcan, *IEEE J. Sel. Top. Quantum Electron.* **2020**, *26*, 1.
[27] D. Mengu, Y. Rivenson, A. Ozcan, *ACS Photonics* **2021**, *8*, 324.
[28] D. Mengu, Y. Zhao, N. T. Yardimci, Y. Rivenson, M. Jarrahi, A. Ozcan, *Nanophotonics* **2020**, *1*, DOI 10.1515/nanoph-2020-0291.
[29] M. S. S. Rahman, J. Li, D. Mengu, Y. Rivenson, A. Ozcan, *Light Sci. Appl.* **2021**, *10*, 14.
[30] H. Chen, J. Feng, M. Jiang, Y. Wang, J. Lin, J. Tan, P. Jin, *Engineering* **2021**, *7*, 1483.
[31] C. Liu, Q. Ma, Z. J. Luo, Q. R. Hong, Q. Xiao, H. C. Zhang, L. Miao, W. M. Yu, Q. Cheng, L. Li, T. J. Cui, *Nat. Electron.* **2022**, *5*, 113.
[32] D. Mengu, A. Ozcan, *Adv. Opt. Mater.* **2022**, *10*, 2200281.
[33] Y. Li, Y. Luo, D. Mengu, B. Bai, A. Ozcan, *Light Adv. Manuf.* **2023**, *4*, 1.
[34] C.-Y. Shen, J. Li, D. Mengu, A. Ozcan, *Adv. Intell. Syst.* **n.d.**, *n/a*, 2300300.
[35] Y. Luo, Y. Zhao, J. Li, E. Cetintas, Y. Rivenson, M. Jarrahi, A. Ozcan, *ArXiv210706586 Phys.* **2021**.
[36] B. Bai, Y. Li, Y. Luo, X. Li, E. Çetintaş, M. Jarrahi, A. Ozcan, *Light Sci. Appl.* **2023**, *12*, 69.
[37] J. Li, D. Mengu, N. T. Yardimci, Y. Luo, X. Li, M. Veli, Y. Rivenson, M. Jarrahi, A. Ozcan, *Sci. Adv.* **2021**, *7*, eabd7690.
[38] "Snapshot multispectral imaging using a diffractive optical network | Light: Science & Applications," can be found under https://www.nature.com/articles/s41377-023-01135-0, **n.d.**
[39] "Super-resolution image display using diffractive decoders | Science Advances," can be found under https://www.science.org/doi/10.1126/sciadv.add3433, **n.d.**
[40] C. Qian, X. Lin, X. Lin, J. Xu, Y. Sun, E. Li, B. Zhang, H. Chen, *Light Sci. Appl.* **2020**, *9*, 1.

[41] Y. Luo, D. Mengu, A. Ozcan, **n.d.**, 24.
[42] B. Bai, Y. Luo, T. Gan, J. Hu, Y. Li, Y. Zhao, D. Mengu, M. Jarrahi, A. Ozcan, *eLight* **2022**, *2*, 14.
[43] J. Li, T. Gan, Y. Zhao, B. Bai, C.-Y. Shen, S. Sun, M. Jarrahi, A. Ozcan, *Sci. Adv.* **2023**, *9*, eadg1505.
[44] B. Bai, X. Yang, T. Gan, J. Li, D. Mengu, M. Jarrahi, A. Ozcan, **2023**, DOI 10.48550/arXiv.2308.15019.
[45] O. Kulce, D. Mengu, Y. Rivenson, A. Ozcan, *ArXiv210809833 Phys.* **2021**.
[46] J. Li, T. Gan, B. Bai, Y. Luo, M. Jarrahi, A. Ozcan, *Adv. Photonics* **2023**, *5*, 016003.
[47] J. Li, Y.-C. Hung, O. Kulce, D. Mengu, A. Ozcan, *Light Sci. Appl.* **2022**, *11*, 153.
[48] Y. Li, T. Gan, B. Bai, Ç. Işıl, M. Jarrahi, A. Ozcan, *Adv. Photonics* **2023**, *5*, 046009.
[49] D. P. Kingma, J. Ba, *ArXiv14126980 Cs* **2014**.
[50] A. Paszke, S. Gross, F. Massa, A. Lerer, J. Bradbury, G. Chanan, T. Killeen, Z. Lin, N. Gimelshein, L. Antiga, A. Desmaison, A. Köpf, E. Yang, Z. DeVito, M. Raison, A. Tejani, S. Chilamkurthy, B. Steiner, L. Fang, J. Bai, S. Chintala, *ArXiv191201703 Cs Stat* **2019**.
[51] O. Kulce, D. Mengu, Y. Rivenson, A. Ozcan, *Light Sci. Appl.* **2021**, *10*, 25.

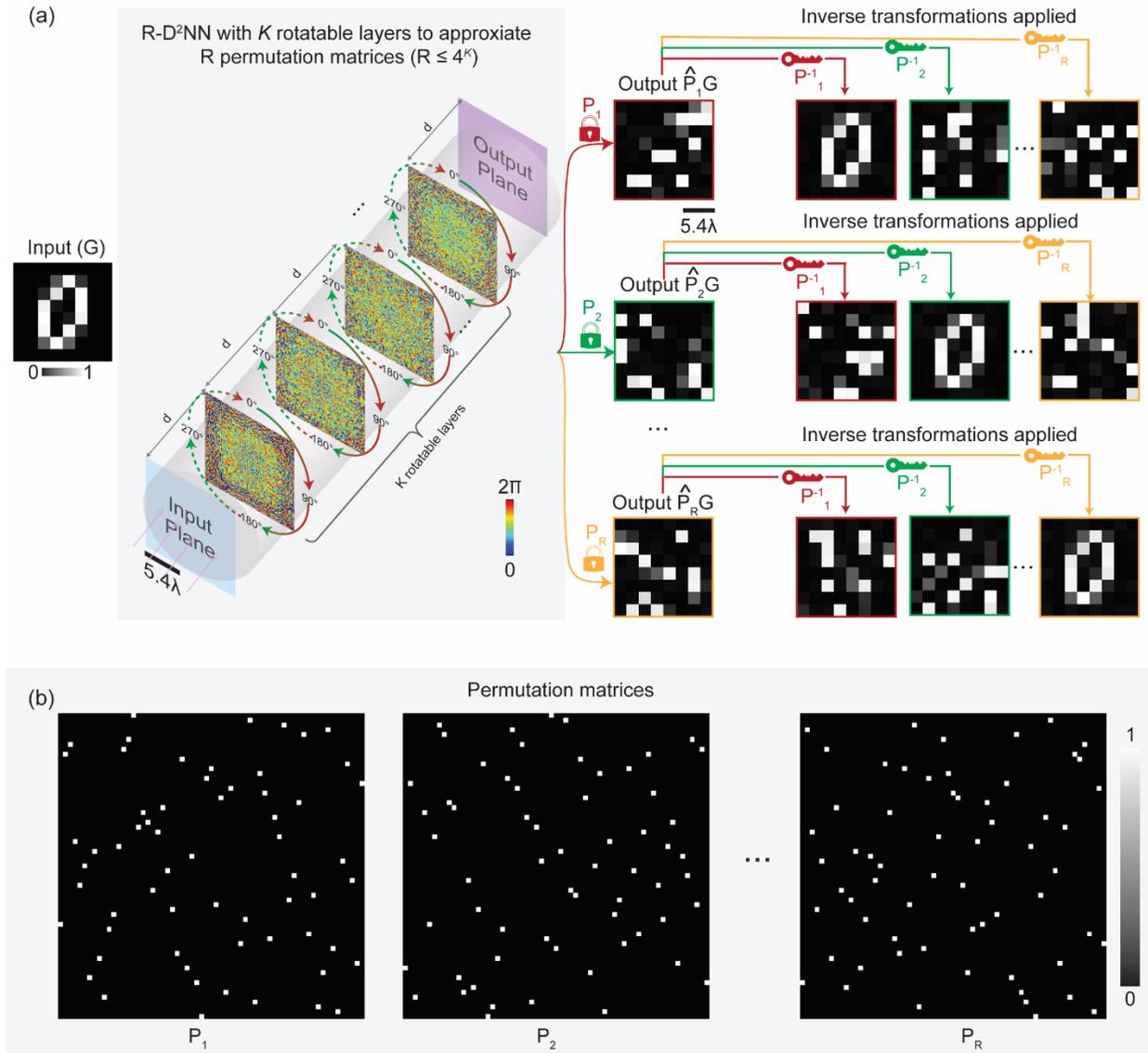

**Figure 1 Schematic of a reconfigurable multiplexed diffractive deep neural network (R-D$^2$NN) with $K$ rotatable layers.** (a) The R-D$^2$NN encompasses $K$ rotatable diffractive layers, each of which permits four independent orientations: $0°$, $90°$, $180°$, and $270°$. Each unique combination of the entire $K$ rotatable layers corresponds to a rotation state of the R-D$^2$NN. Therefore, a $K$-layer R-D$^2$NN can execute $4^K$ distinct rotation states, performing up to $4^K$ unique transformations - one assigned to each rotation state. The presented $K$-layer rotatable R-D$^2$NN is trained to approximate $R$ unique permutation operations with $R$ distinct rotation states ($R \leq 4^K$), ultimately resulting in different permutated/encrypted images at the

output plane with different rotation states. The decryption is implemented by applying the inverse of the permutation matrix. Only applying the specific inverse permutation operation enables accurate image recovery/decryption, while applying other mismatched permutation operations results in speckle-like images. (b) Preassigned permutation matrices: $P_1, P_2, \ldots P_R$.

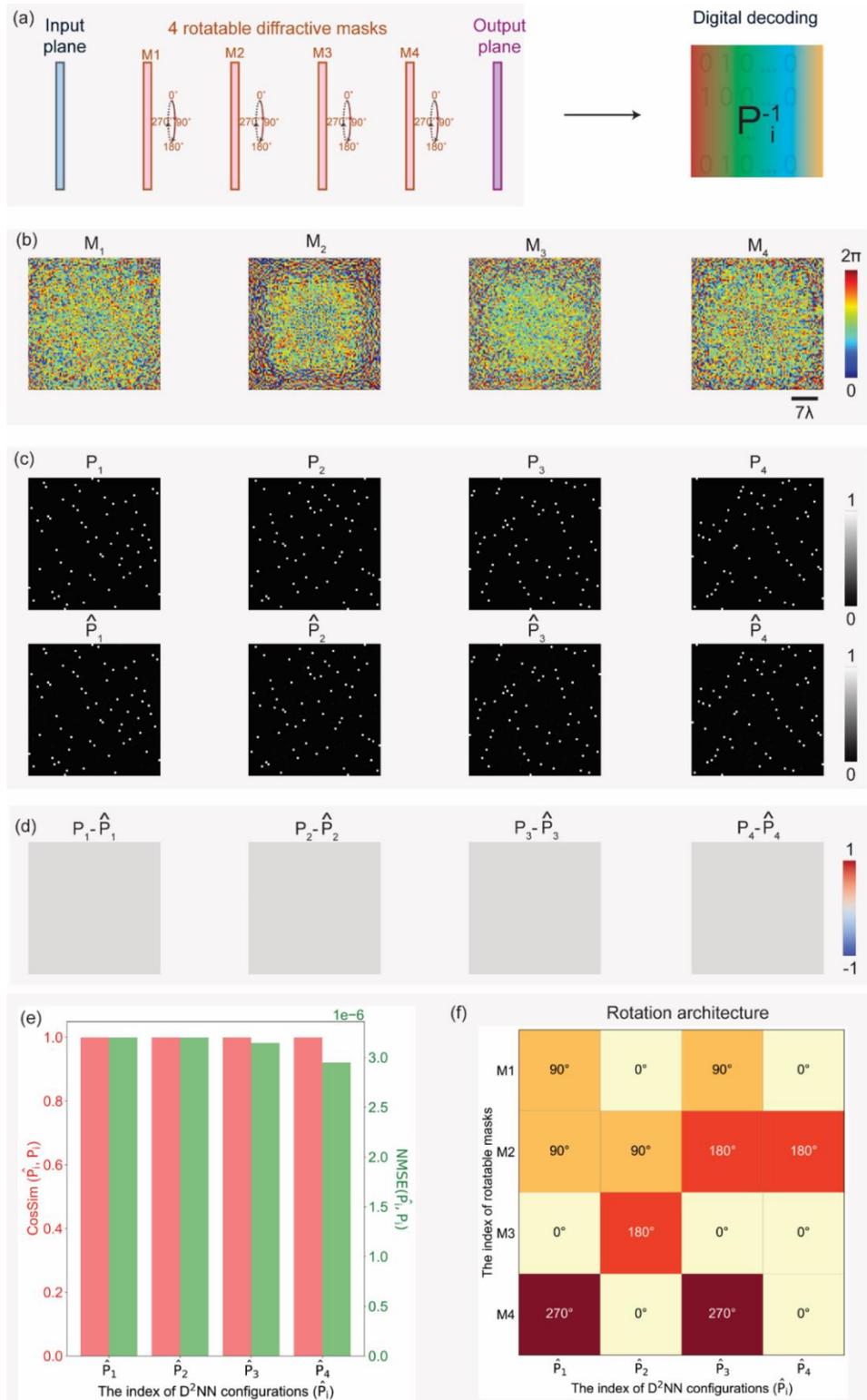

**Figure 2 Example of a 4-layer R-D$^2$NN design.** (a) The physical layout depicts a 4-layer rotatable R-D$^2$NN, approximating four permutation matrices between the input and output FOVs with four rotation states.

(b) Optimized diffractive layers of the rotatable R-D²NN design. There are $1.5RN_iN_o$ diffractive features/neurons evenly distributed on four diffractive layers ($R = 4, N_i = N_o = 64$). (c) The target permutation matrices ($P_i, i = 1, 2, 3, 4$) and their all-optical versions ($\hat{P}_i, i = 1, 2, 3, 4$) executed by the R-D²NN. (d) The difference between the target and the all-optically executed permutation matrices. (e) The cosine similarity and permutation error between the desired and the estimated permutation matrices. (f) The randomly selected rotation architectures for each R-D²NN configuration/rotation state.

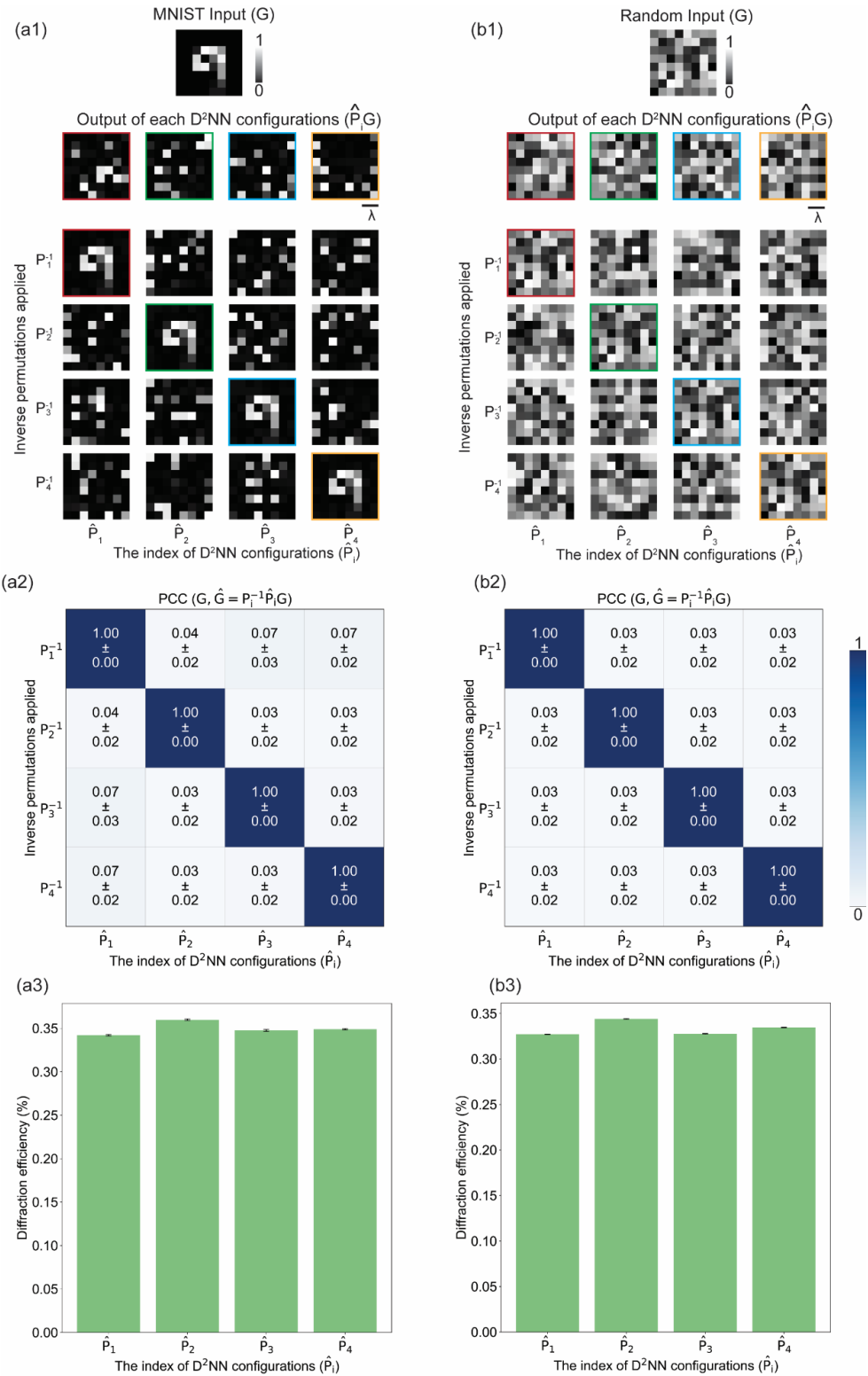

**Figure 3 Blind testing results of the R-D$^2$NN model shown in Figure 2 with different datasets.** (a1), (b1) An example of the input intensity image, all-optical encryption results, and digital decryption results with a handwritten MNIST digit and a randomly generated input image. (a2), (b2) The mean PCC values and the corresponding standard deviation values between the ground truth images ($G$) and the recovered images ($\hat{G} = P_i^{-1}\hat{P}_i G$) for the entire 20,000 blind testing results with the handwritten MNIST dataset and a randomly generated dataset. (a3), (b3) The diffraction efficiency of all 20,000 blind testing results with handwritten MNIST dataset and customized random dataset. The standard deviations (represented with the error bars) in (a3) and (b3) are calculated across the entire 20,000 testing dataset.

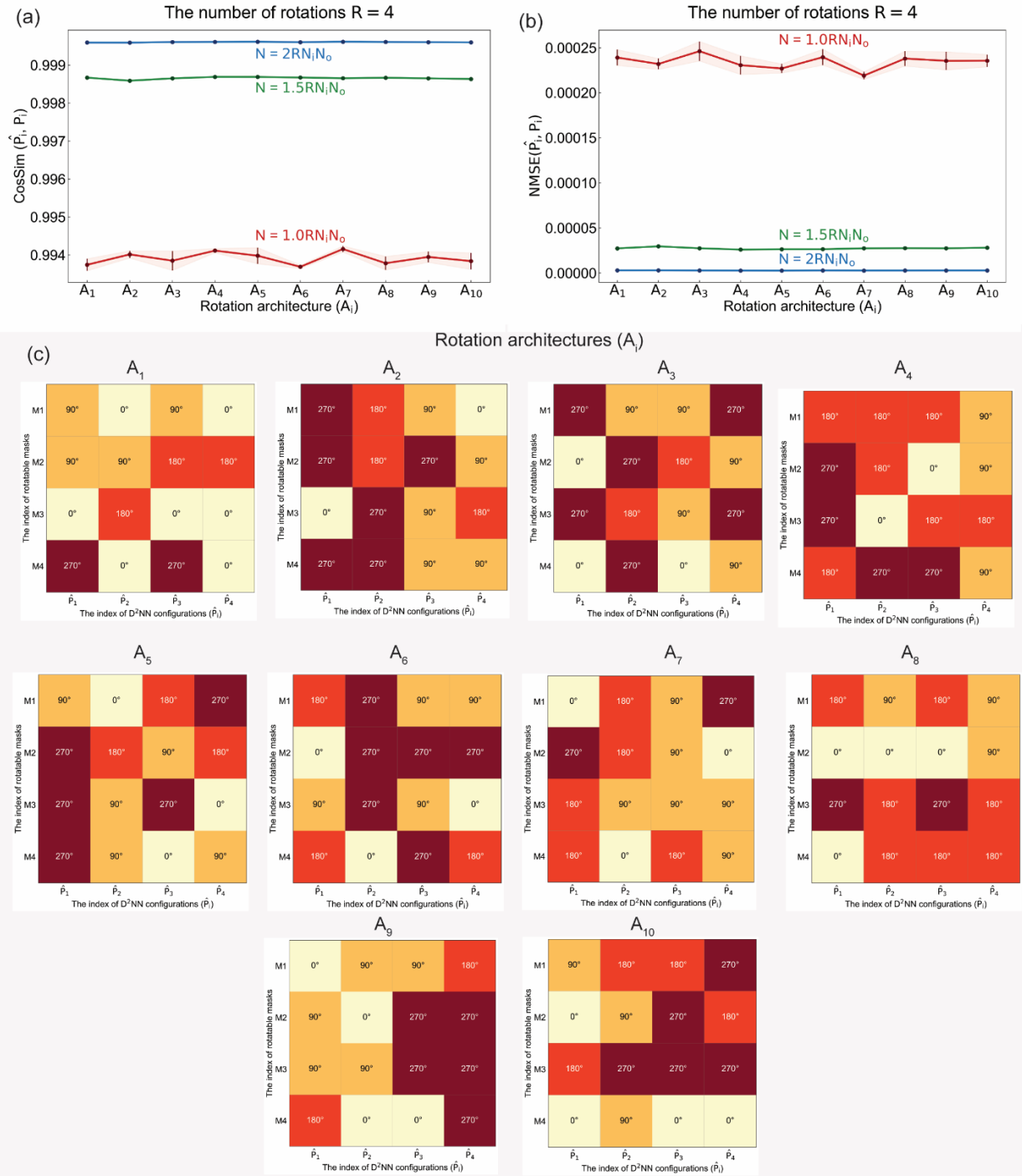

**Figure 4 The impact of different rotation architectures on the performance of a 4-layer R-$D^2$NN**. (a) The cosine similarity, (b) the permutation error, between the desired permutation matrices ($P_i, i = 1, 2, 3, 4$) and their estimated versions ($\hat{P}_i, i = 1, 2, 3, 4$) as a function of ten different rotation architectures ($A_i, i =$

$1, \ldots, 10$) and $N$. The standard deviations (error bars) in (a) and (b) are calculated across 4 estimated permutation matrices. (c) Ten rotation architectures for each R-D$^2$NN design.

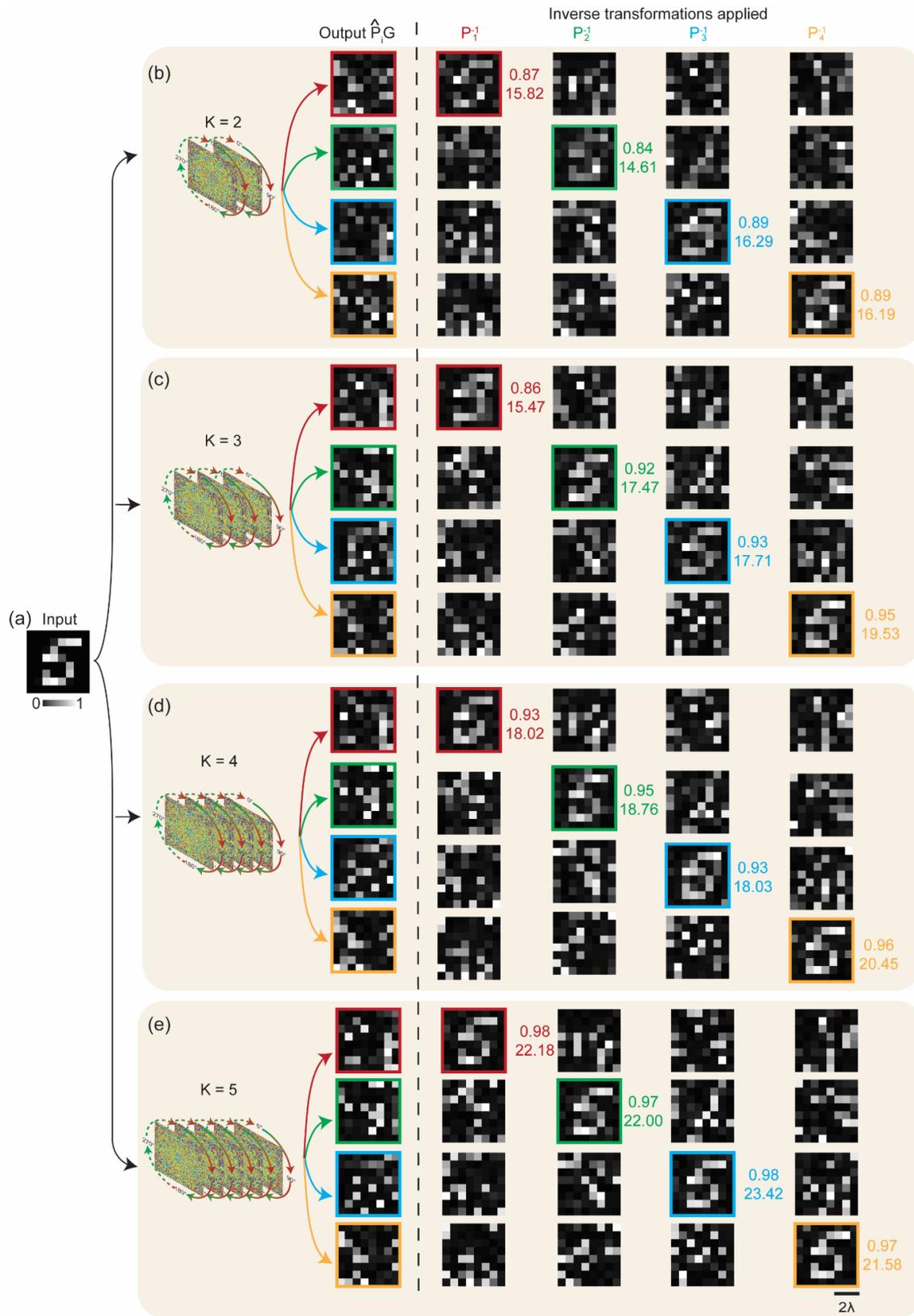

**Figure 5 Performance of $R$-$D^2NN$ with varying numbers of rotatable diffractive layers while maintaining**

**the same number (*N*) of diffractive features.** (a) The input intensity image. (b), (c), (d), (e) The reconfigurable diffractive designs keep the same number of diffractive features $N = 1.0RN_iN_o$ ($R = 4, N_i = N_o = 64$) for different numbers of rotatable layers (2, 3, 4, and 5). PCC (up) and PSNR (down) values between the decrypted and the original images are shown on the right side of each diagonal image.

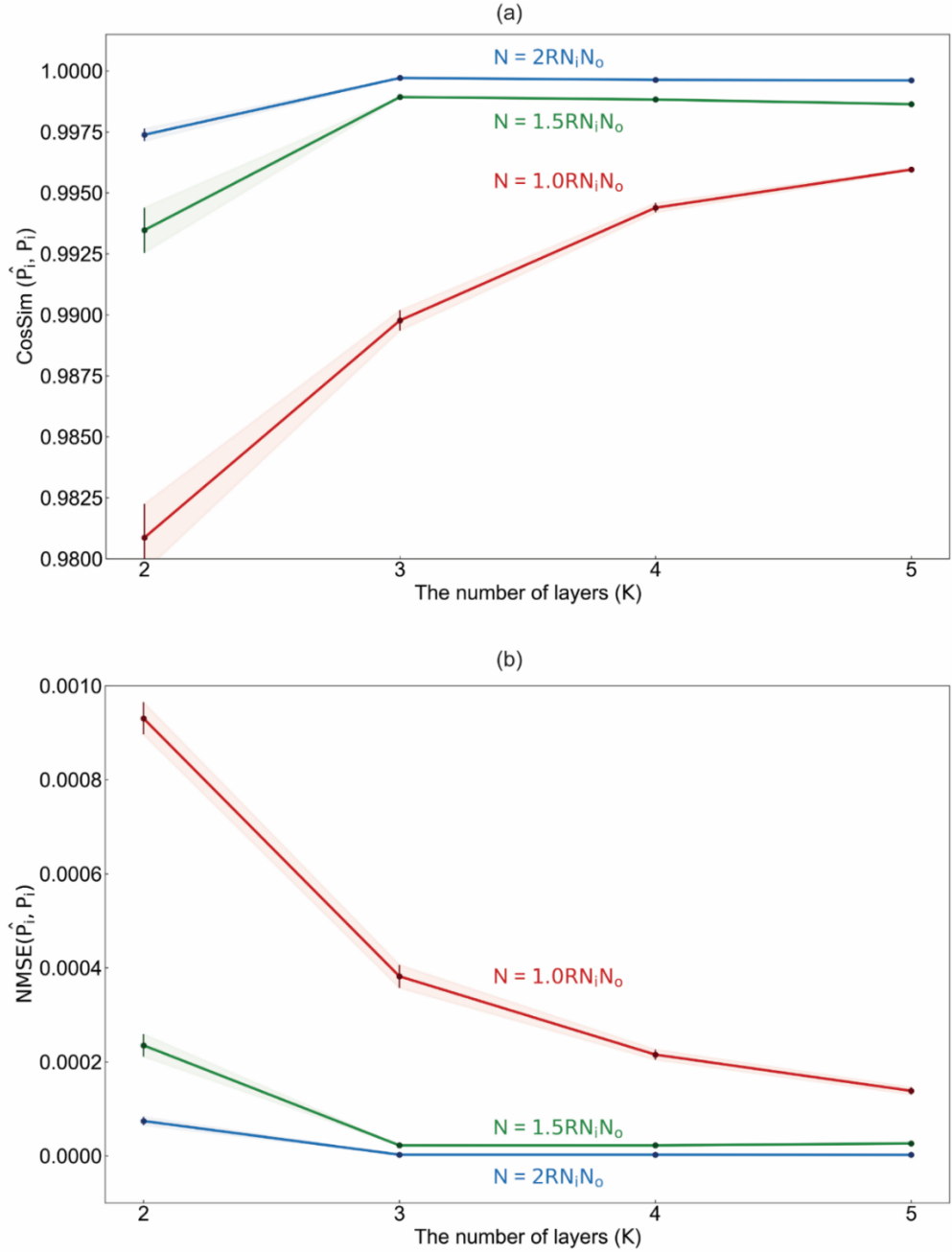

**Figure. 6 Impact of the distribution of the diffractive features among rotatable diffractive layers on the performance of R-D$^2$NN**. (a) The cosine similarity, (b) the permutation error, between the desired permutation matrices ($P_i, i = 1, 2, 3, 4$) and their all-optical versions ($\hat{P}_i, i = 1, 2, 3, 4$) for different combinations of $K$ and $N$. The standard deviations (error bars) of these metrics are calculated across 4 estimated permutation matrices.

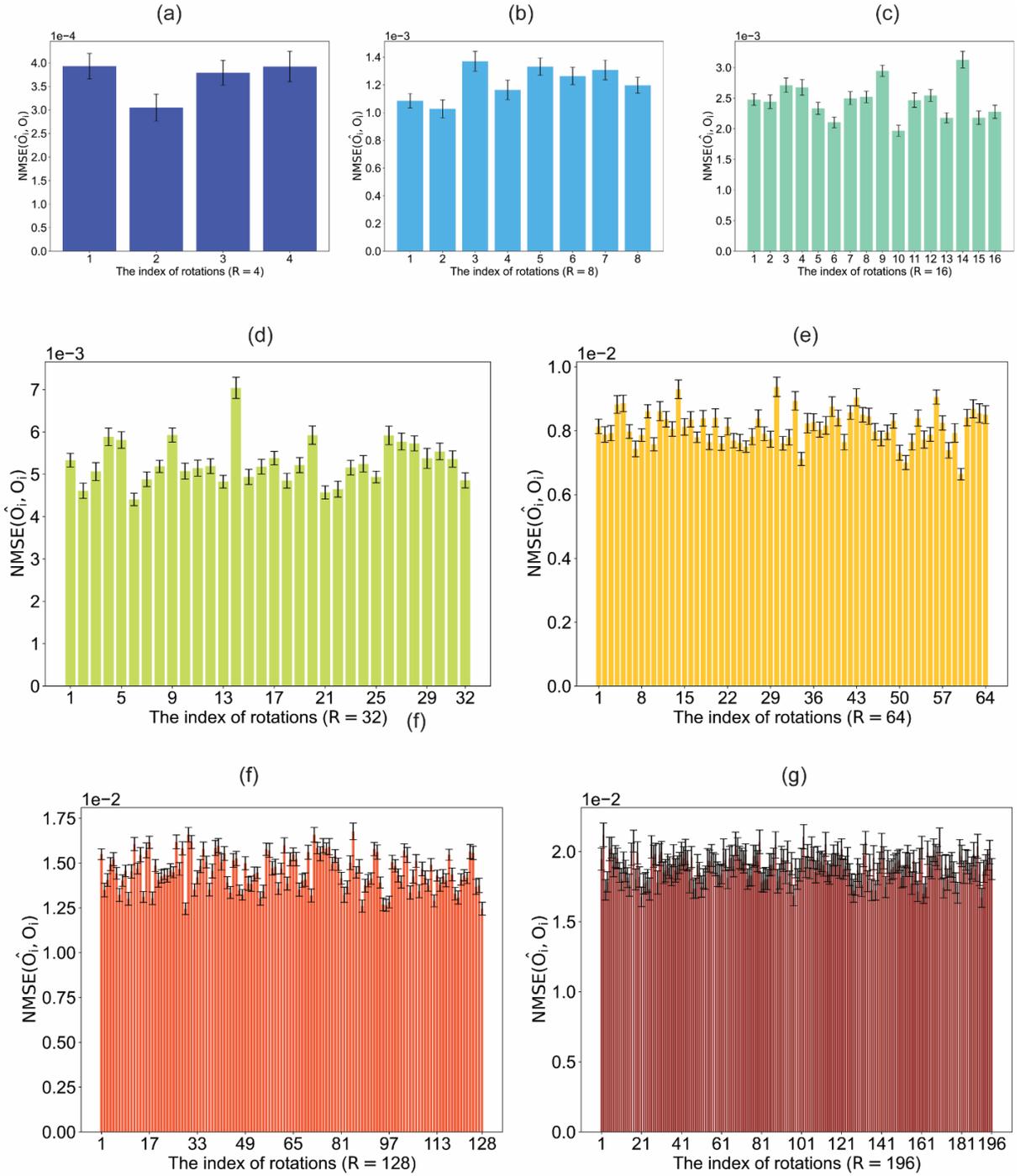

**Figure 7 All-optical permutation performance as a function of $R$ in a four-layer $R$-$D^2$NN design with $N = 1.5RN_iN_o$ ($N_i = N_o = 64$).** (a) $R = 4$; (b) $R = 8$; (c) $R = 16$; (d) $R = 32$; (e) $R = 64$; (f) $R = 128$; (g) $R = 196$. The standard deviations are calculated across the entire 10,000 test dataset.

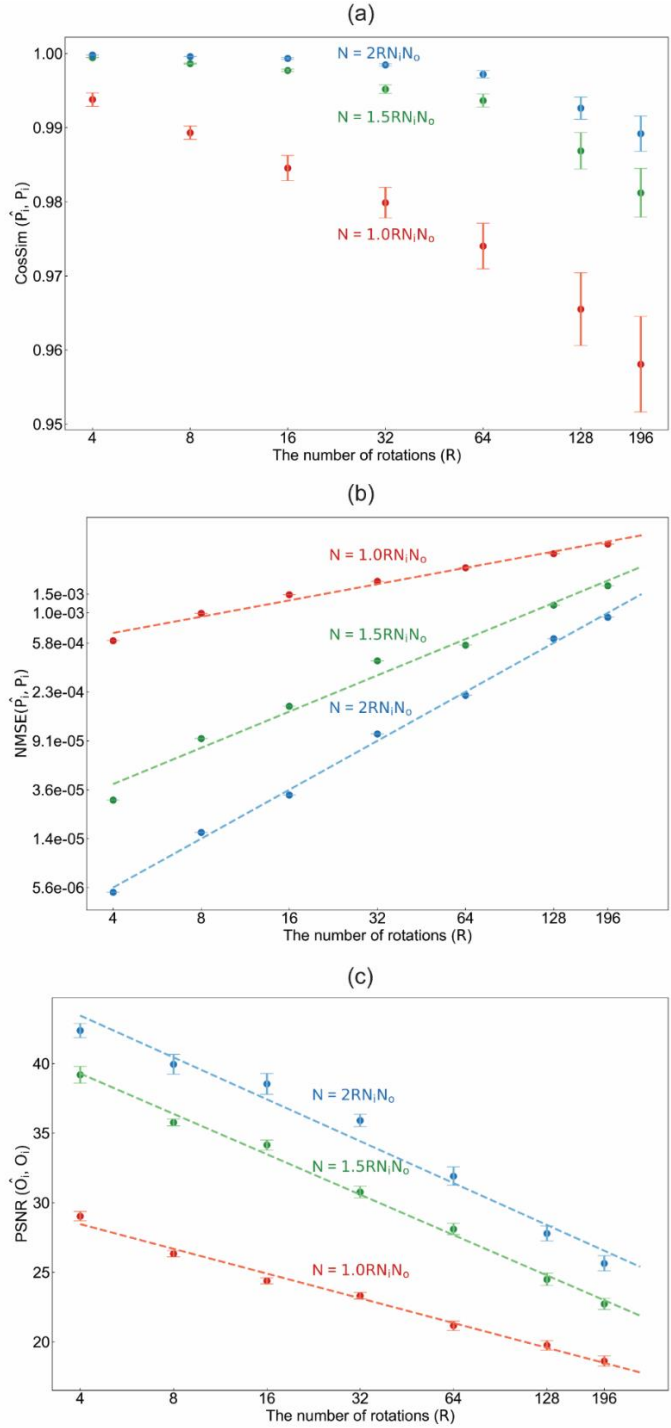

**Figure. 8 Varying number of rotation states for a 4-layer R-D$^2$NN design.** (a) The cosine similarity, (b) the permutation error, between the desired permutation matrices ($P_i$) and their estimated versions ($\hat{P}_i$) for different combinations of $R$ and $N$. (c) The PSNR of 10,000 blind testing results with a randomly generated

dataset. $N_i = N_o = 25$. The standard deviations (error bars) in (a) and (b) are calculated across $R$ permutation matrices, while the standard deviations (error bars) in (c) are calculated across the entire 10,000 random testing dataset.

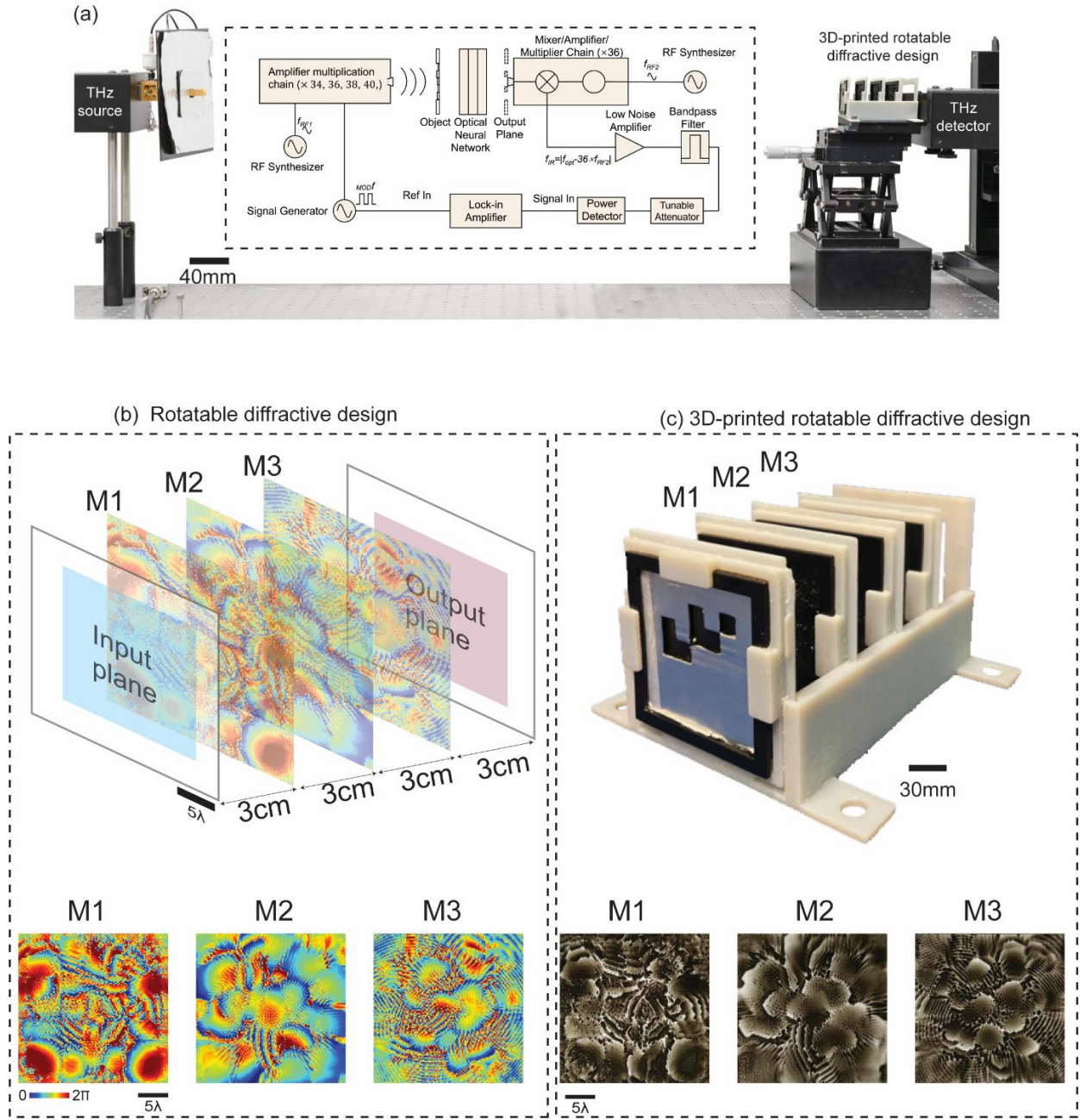

**Figure 9 Experimental setup.** (a) Schematic of the THz experimental setup. (b) Layout of the vaccinated rotatable diffractive design. In this design, M1 and M2 are fixed, while M3 is rotated $0°$ or $180°$. These two

rotation configurations are used to approximate two permutation matrices $P_1$ and $P_2$. (c) Fabricated three-layer rotatable R-D$^2$NN.

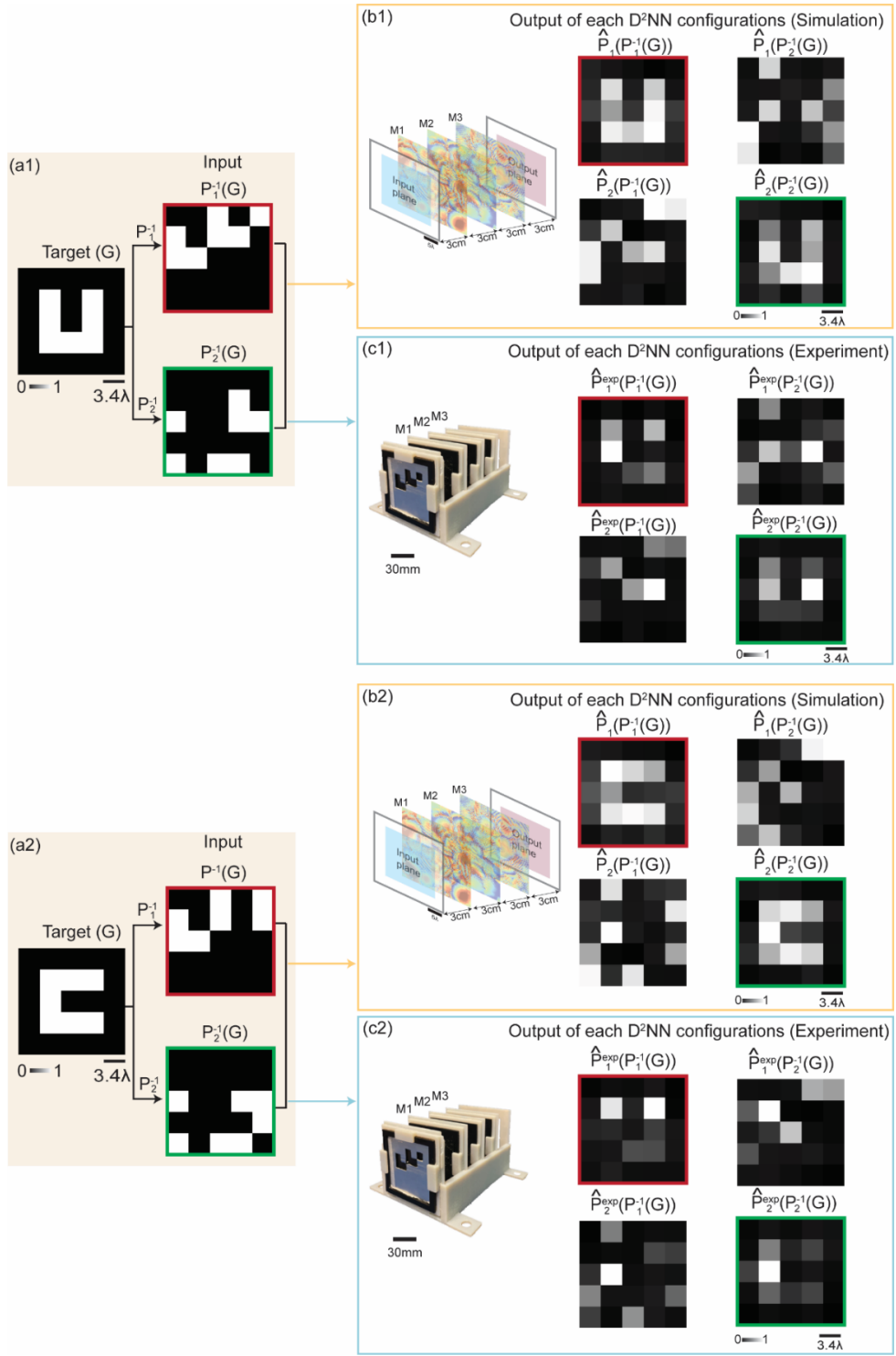

**Figure 10 Experimental results of R-$D^2$NN.** (a1), (a2) Target images and their corresponding versions permuted by the inverse permutation matrices $P_1^{-1}$, and $P_2^{-1}$. (b1), (b2) The vaccinated R-$D^2$NN design and the corresponding simulation results. (c1), (c2) Experimental results.